\documentclass[twocolumn,showpacs,preprintnumbers,amsmath,amssymb,superscriptaddress]{revtex4}
\usepackage{dcolumn}
\usepackage{amsmath}
\usepackage{graphicx}
\usepackage{latexsym}
\usepackage{amsfonts}
\usepackage{amssymb}
\usepackage{color}
\DeclareGraphicsExtensions{.pdf,.gif,.jpg}
\newcommand{\be}{\begin{equation}}
\newcommand{\ee}{\end{equation}}
\newcommand{\beq}{\begin{eqnarray}}
\newcommand{\eeq}{\end{eqnarray}}

\begin{document}

\newcommand\xmas{\mbox{$x_{\rm mas}$}}
\newcommand\rmas{\mbox{$r_{\rm mas}$}}
\newcommand\Xmas{\mbox{$X_{\rm mas}$}}
\newcommand\Hinf{\mbox{${\rm H}_{\rm inf}$}}
\newcommand\Omegainf{\mbox{$\Omega_{\rm inf}$}}
\newcommand\Omegabg{\mbox{$\Omega_{\rm bg}$}}
\newcommand\Hbg{\mbox{${\rm H}_{\rm bg}$}}
\newcommand\rhobg{\mbox{$\rho_{\rm bg}$}}
\newcommand\rhorbg{\mbox{$\rho^{\rm bg}_{\rm r}$}}
\newcommand\rhoinf{\mbox{$\rho_{\rm inf}$}}
\newcommand\kinf{\mbox{$k_{\rm inf}$}}
\newcommand\kbg{\mbox{$k_{\rm bg}$}}
\newcommand\ainf{\mbox{$a_{\rm inf}$}}
\newcommand\abg{\mbox{$a_{\rm bg}$}}
\newcommand\xpbg{\mbox{$x_{\rm p}^{\rm bg}$}}
\newcommand\xpinf{\mbox{$x_{\rm p}^{\rm inf}$}}
\newcommand\xmasbg{\mbox{$x_{\rm mas}^{\rm bg}$}}
\newcommand\xmasinf{\mbox{$x_{\rm mas}^{\rm inf}$}}
\newcommand\thetainf{\mbox{$\theta^{\rm inf}$}}
\newcommand\thetabg{\mbox{$\theta^{\rm bg}$}}
\newcommand\xHbg{\mbox{$1/{\rm H}_{\rm bg}$}}
\newcommand\xHinf{\mbox{$1/{\rm H}_{\rm inf}$}}

\title{GW band structure of InAs and GaAs in the wurtzite phase}

\author{Z. Zanolli}
\affiliation{Department of Physics,  Lund University, S\"olvegatan 14 A, 223 62 Lund, Sweden}
\author{F. Fuchs}
\affiliation{Institut f\"ur Festk\"orpertheorie und optik, Friedrich-Schiller-Universit\"at, Max-Wien-Platz 1, 07743 Jena, Germany}
\author{J. Furthm\"uller}
\affiliation{Institut f\"ur Festk\"orpertheorie und optik, Friedrich-Schiller-Universit\"at, Max-Wien-Platz 1, 07743 Jena, Germany}
\author{U. von Barth}
\affiliation{Department of Physics,  Lund University, S\"olvegatan 14 A, 223 62 Lund, Sweden}
\author {F. Bechstedt}
\affiliation{Institut f\"ur Festk\"orpertheorie und optik, Friedrich-Schiller-Universit\"at, Max-Wien-Platz 1, 07743 Jena, Germany}

\date{\today}

\begin{abstract}
We report the first quasiparticle calculations of the newly observed wurtzite polymorph of InAs and GaAs. 
The calculations are performed in the GW approximation using plane waves and pseudopotentials. For comparison we also report the study of the zinc-blende phase within the same approximations.
In the InAs compound the In 4{\it d} electrons play a very important role: whether they are frozen in the core or not, leads either to a correct or a wrong band ordering (negative gap) within the Local Density Appproximation (LDA). We have calculated the GW band structure in both cases. In the first approach, we have estimated the correction to the {\it pd} repulsion calculated within the LDA and included this effect in the calculation of the GW corrections to the LDA spectrum. In the second case, we circumvent the negative gap problem by first using the screened exchange approximation and then calculating the GW corrections starting from the so obtained eigenvalues and eigenfunctions. This approach leads to a more realistic band-structure
and was also used for GaAs. For both InAs and GaAs in the wurtzite phase we predict an increase of the quasiparticle gap with respect to the zinc-blende polytype.
\end{abstract}


\maketitle

\section{Introduction}

With the development of the growth techniques such as Chemical Beam Epitaxy (CBE) new nanostructured materials are being created. Among the technologically important III-V semiconductors, one exceptional example is InAs. Indeed InAs nanowires (NWs) grow purely 
in wurtzite structure with [0001] orientation when InAs (111)B is chosen as substrate \cite{Zanolli2006a}. 
Recalling that the zinc-blende structure ($zb, 3C$, space group $F\overline43m$ ($T^2_d$)) is the stable phase of bulk InAs, the new  hexagonal phase (wurtzite $wz, 2H$, space group $P6_3mc$ ($C^6_{4v}$)) clearly represents a theoretical challenge.
Obviously, the new phase calls for a proper theoretical investigation
 in order to better understand its physical properties and give a correct interpretation of recent experiments conducted on InAs-based NWs.  Some examples are given by the  photoluminescence (PL) measurements performed by one of the authors on InAs-InP core-shell strained NWs \cite{Zanolli2006}, photocurrent measurements 
on InAs NWs with a centrally placed InAs$_{1-x}$P$_x$ segment \cite{Tragardh2006}, and electron g-factor measurements on InAs NW quantum dots 
\cite{Bjork2005}. 
GaAs NWs grown on GaAs (111)B substrate, instead,  typically  consist of alternating wurtzite and zinc-blende segments.

The electronic structures of the wurtzite polytypes of InAs and GaAs are almost unknown. This is certainly true for the variation of the fundamental gap of the polytype holds in particular for the varation of the fundamental gap with the polytype. Moreover, once the full band structure of the new polymorph is obtained, it can be used to extract parameters needed for tight binding and $\bf {k}{\cdot} \bf{p}$ perturbation theory calculations. Furthermore, the so obtained $\bf {k}{\cdot} \bf{p}$ Hamiltonian is a necessary input to electron transport studies on the same material, with obvious consequences to nano-electronics.

\section{Methods of investigation}

Since there are no measured properties of InAs in the wurtzite phase, the choice of an {\it ab initio} method is highly appropriate because it does not require any previous knowledge of the material parameters and allows one to predict numerous properties ranging from the equilibrium lattice constant to the band structure.
The ground state electronic structure of solids is well described by Density Functional Theory (DFT) \cite{Kohn1964_DFT} in the Local Density Approximation (LDA) \cite{Kohn-Sham1965}. Hence, these calculations are specifically suited for structural studies. However, the LDA is not appropriate for describing electronic excitations. In particular, the calculated band gaps are underestimated. The excited-state properties of a many-electron system, such as the band structure, require a quasiparticle calculation in order to properly account for many-body effects. These calculations rely on the concept of the single-particle Green function G, whose exact determination requires the complete knowledge of the electronic self-energy ${\Sigma}$. The latter is a non-Hermitian, non-local, and energy-dependent operator which describes exchange and correlation effects beyond the Hartree approximation and which can only be calculated approximately. A useful approximation is the dynamically screened exchange or GW approximation (GWA) \cite{Hedin1965, Hedin1969}. 
Since the electron density is accurately given by the LDA, the wave functions of this approximation are usually rather close to the quasiparticle wave functions obtained from the self-energy operator. Therefore, the quasiparticle band structure is usually calculated perturbatively starting from the LDA wave functions and eigenenergies.

Our goal is the GW band structure of InAs and GaAs in the wurtzite phase.  In the case of InAs, the main obstacle on this path is the fact that the correct procedure to include 
 the In 4{\it d} electrons among the valence states will erroneously predict InAs to be a metal within the LDA.  Hence, the LDA eigenfunctions and energies are no longer a good starting point for the GW calculation. 
The main reason for the failure of the LDA  is an overestimation of the {\it pd} coupling as has been concluded by several authors \cite{Rohlfing1995, Bachelet1985, Christensen1988, Wei1988} in different materials like, for instance, InN \cite{Furth2005}, GaAs \cite{Tiago2004} and II-VI compounds \cite{Zakharov1994}.
If, instead, the In 4{\it d} states are frozen into the core, the correct band ordering can be re-established, making the LDA results a reasonable starting point for a subsequent GW calculation.  For this specific reason, in the first part of this study, we have used In pseudopotentials for which the {\it d} electrons are treated as core states and their self-interaction corrections  (SICs) are included  in the underlying free-atom calculation \cite{Rieger1995}.

Another way of circumventing the negative gap problem, is to perform a calculation using an approximate self-energy which restores the correct band ordering. The screened-exchange approximation is such a self-energy. The GW calculation is then based on the latter results. The {\it pd}-repulsion is correctly taken into account.

Even if there is no problem with the band ordering, as is the case in GaAs, it could still be valuable to base the GW calculation on an electronic structure obtained from a self-energy 
producing results much closer to the GWA. This would lead to final results closer to a self-consistent GW calculation. We have found this to be important also in GaAs. Previous calculations  treating the GW self-energy as a perturbation to the LDA potential have resulted a too small band gap. This is true also in  all-electron calculations \cite{Arnaud2000}. Thus, it appears that in materials with $d$-electrons like, for instance, InAs and GaAs, self-consistency is an important issue.


Because of the novelty of the wurtzite polymorph, there are few theoretical studies at the DFT/LDA level \cite{Wang2002, Wang2003} reported in the literature and no GW calculation. Besides, it should also be noted that these DFT/LDA calculations were done assuming the In 4{\it d} electrons frozen in the core and this, as we will discuss in this article, leads to an underestimation of the equilibrium lattice parameters  and to a misleading description of the band-structure if the {\it pd} repulsion is not properly taken into account.

On the basis of total-energy calculations within DFT  it has been concluded  \cite{Crain1994} that the wurtzite phase is a metastable high-pressure modification of the InAs and GaAs compounds. The mechanism behind this is a first-order phase transition induced by pressure.
From LDA calculations it has been demonstrated \cite{Yeh1994} that in the case of zinc-blende compounds with a direct conduction band minimum at $\Gamma_{1c}$ and the $L_{1c}$ state above the $\Gamma_{1c}$, the corresponding wurtzite compound will also have a direct gap which is  slightly larger. 

In this paper, we present the results of {\it ab initio} DFT/LDA  quasiparticle calculations of band structures for both zinc-blende and wurtzite phases of InAs with and without the inclusion of the In 4{\it d} electrons among the valence states. The paper is organized as follows. At first we present the technical details of the calculations.
Then we investigate the role played by the In 4{\it d} electrons by comparing  the US-PP LDA band structures calculated with and without the inclusion of the {\it d} states in the valence band.
By means of a further comparison with the experimental ({\it zb}) energy of the {\it d} states we have estimated a correction factor for the band structure when calculated with the {\it d} states frozen in the core. After obtaining the GW band structure in this case, we have applied this correction to arrive at the final wurtzite band-structure. 
Finally we treat the case with the {\it d} states in the valence band by using the screened exchange (SX) functional for exchange and correlation
\cite{Hedin1965}.
This gives a  band structure with a correct band ordering and hence it is a valid starting point for the evaluation of the GW corrections.

 \section{Computational details}

All the calculations are performed with the VASP code \cite{VASP},
a software package for {\it ab initio} simulations which relies on the description of the electron-ion interaction via ultra-soft non-norm-conserving 
pseudopotentials (US-PP) \cite{Vanderbilt1985} or the projector-augmented wave (PAW) method \cite{Bloechl1994, Kresse1999} and a plane-wave expansion of the eigenfunctions.  Both kinds (US and PAW) of pseudopotentials are used throughout this study, for both polymorphs and for both treatments of the In 4{\it d}  states. Hence, the number of studied cases adds up to 8, for each of which an accurate convergence study was performed, in such a way as to achieve an absolute convergence of a few meV (always better than 10 meV) for every electronic level. 
In this paper we use a simplification for the self-energy based on a model dielectric function which, in turn, is based on a plasmon pole approximation
for the dynamics \cite{Bechstedt1992, Furthmuller2002}.

We have performed the convergence study for the two polymorphs
with the In 4{\it d} electrons as either valence ({\it d}-val) or core ({\it d}-core) states. For the zinc-blende phase we have used a $\Gamma$-centered Monkhorst-Pack \cite{Monk1976} grid with a $11 \times 11 \times 11$ mesh in reciprocal space. For the wurtzite phase, instead, a $8  \times  8  \times  8$ mesh was used.

As kinetic energy cutoffs we used 262 eV  
when the {\it d} states are in the valence band (for both polymorphs and both pseudopotentials). When the {\it d} electrons are included in the core we used a cutoff 222 eV  
and 202 eV 
for the {\it zb} and {\it wz} cases, respectively.

\begin{figure}[thb]
\centering
\includegraphics[width=8cm]{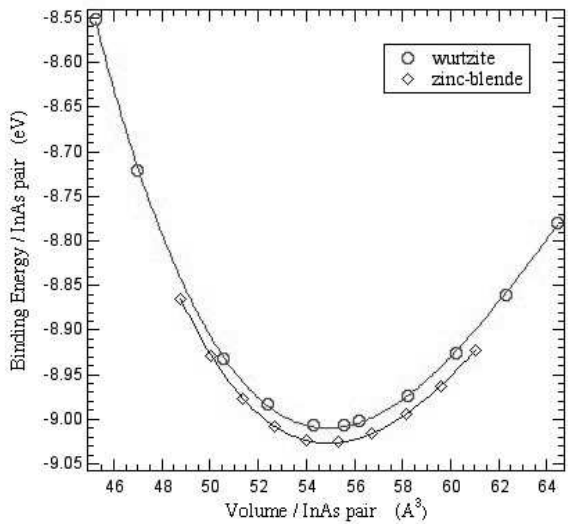} 
\caption{The normalized total energy versus volume for one InAs pair for the zinc-blende (diamonds) and the wurtzite (circles) polymorphs. Total energies are calculated by using US pseudopotentials.}
\label{fig:Energy-Vol} 
\end{figure}

The structure optimization has been performed by minimization of the total energy.  In the wurtzite case we have assumed the ideal values for the $c/a$ ratio and the internal parameter ($u$). 
The resulting  energy versus volume curves, normalized to one InAs pair, are displayed in Fig. \ref{fig:Energy-Vol}. Here, we can see that the {\it zb} is correctly predicted to be the stable phase, since its binding energy is more negative than the {\it wz} by $\sim 18$ meV. 

The results obtained from the structure optimization of InAs for both zinc-blende and wurtzite phases are collected in Table \ref{tab:latt_const} together with 
experimental results for lattice constants. The experimental values for wurtzite are taken from TEM measurements \cite{Zanolli2006a}, while, for the zinc-blende, the experimental parameter measured at room temperature
\cite{Landolt1982} has been extrapolated to zero degrees Kelvin.
From this comparison we see that the calculated lattice constants when the {\it d} electrons are treated as valence states are very close to experimental results. Hence, these values have been used throughout our further investigation. 

The results reported in literature for the {\it wz} polymorph are obtained using HGH pseudopotentials in the {\it d}-core approximation \cite{Wang2002}: $a = 4.192$~\AA, $u = 0.3755$, $c = 6.844$~\AA.
In the same article and with the same approach is reported $a = 5.921$~{\AA} for the {\it zb}. 
More results obtained at the LDA level with pseudopotentials ({\it d}-core) are reported for the {\it zb} phase as, for instance,  $a = 5.04$~{\AA} \cite{Massidda1990}, $a = 5.902 $~{\AA} \cite{VanCamp1990}, $a = 5.95$~{\AA} \cite{Boguslawski1989}, $a = 5.906$~{\AA} \cite{Zhang1987}.

\begin{table}[hbt]
\caption{Lattice constants (\AA) for InAs in the zinc-blende and wurtzite phases. US denotes Ultra Soft pseudopotentials with the In 4{\it d} electrons included among the valence ({\it d}-val) or core ({\it d}-core) states.}
    \begin{center}
       \begin{tabular}{c|c|c|c}
           \hline \hline
         	                                    &    $a_{zb}$    &     $a_{wz}$      &   $c_{wz}$     \\ 
        	   \hline
           experimental           & 6.0542 {\AA} & 4.2839   {\AA} &  6.9954 {\AA} \\
           US {\it d}-val     & 6.0329 {\AA} & 4.2663   {\AA} &  6.9669 {\AA} \\
           US {\it d}-core   & 5.8023 {\AA} & 4.1060   {\AA} &  6.7051 {\AA} \\
           \hline \hline
       \end{tabular} 
    \end{center}
\label{tab:latt_const}
\end{table}


\section{Band structures}
 \subsection{Kohn-Sham and quasiparticle band structures.}

\begin{figure}[htb]
\centering
\includegraphics[width=8cm]{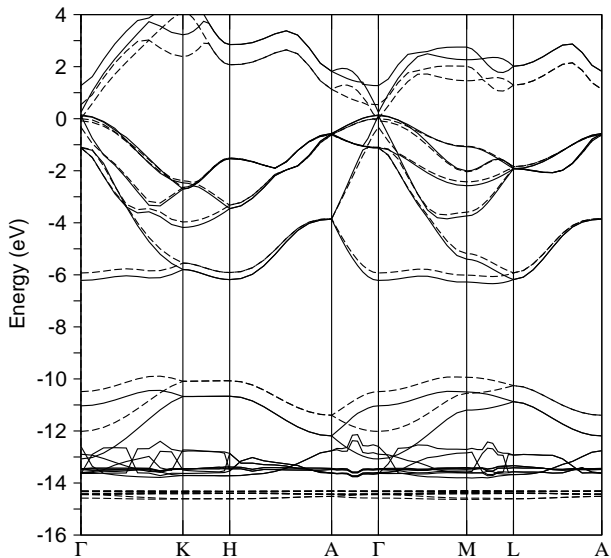}
\caption{Quasiparticle (solid lines) and Kohn-Sham (dashed lines) band structures of InAs wurtzite with the In 4{\it d} included in the valence states: at $\Gamma$ a ``negative gap'' is predicted. }
\label{fig:GW_US_wz_val} 
\end{figure}

The LDA band structures  along high-symmetry lines in the Brillouin-Zone (BZ) were obtained by solving the Kohn-Sham equation \cite{Kohn-Sham1965} at the equilibrium lattice parameters and using US-PP. For the {\it wz}-InAs, these are shown as continuous lines in Fig.~\ref{fig:GW_US_wz_val} and \ref{fig:GW_US_wz_core}  in the {\it d}-val and {\it d}-core case, respectively.
When the In 4{\it d} electrons are included in the valence band, the calculated band structure reveals  problems for both polymorphs: 
a negative {\it sp} gap is obtained, resulting in a ``wrong band ordering''.  The $s$-like $\Gamma_{1c}$ state  - forming normally the conduction band minimum (CBM) - lies below the $p$-like $\Gamma_{15v}$ (or $\Gamma_{6v}$ and $\Gamma_{1v}$) state - constituting the valence band maximum (VBM) in other zinc-blende (or wurtzite) semiconductors. We find $\Gamma_{1c} -  \Gamma_{15v} = - 0.346$~eV  for the zinc-blende  and 
$\Gamma_{1c} - \Gamma_{6v}  = - 0.317$~eV for the wurtzite energy gaps.
The only difference between the two InAs phases is that the  {\it wz} exhibits  crystal field splitting between the $\Gamma_{6v}$ and  $\Gamma_{1v}$ valence band levels.
The calculated value is
$\Delta_{cr} =   \epsilon (\Gamma_{6v}) - \epsilon(\Gamma_{1v}) = 81$~meV.


The  ``negative gap'' of the InAs in the {\it zb} phase has already been observed in other LDA calculations that include the {\it d} states in the valence band as, for instance,  reported in Ref. \cite{Zhu1991}.
%
The main cause of this problem is the known limitation of the DFT-LDA to correctly describe excited states properties like  the band structure. Actually, the latter is properly described within many-body perturbation theory, which requires  a zero:th order Hamiltonian for the perturbation series. This is usually chosen to be the LDA eigenspectrum. 
Unfortunately, in the present case, the wrong starting Hamiltonian, will lead to unreliable GW corrections which actually also results in an erroneous band structure as can be seen from Fig.~\ref{fig:GW_US_wz_val}. 

When the experimental band-gap is as small as it is in the present case  ($E_{gap} = 0.415$ eV at 0 K \cite{Fang1990}) sources of error other than the fundamental one become important like the LDA itself. The LDA becomes less accurate in more inhomogeneous materials such as the III-IV semiconductors and even worse in the II-VI semiconductors with a higher ionicity. Besides, as it will be discussed in the next section, the underestimation of the binding enery of the In $4d$ electrons causes an overestimation of the {\it pd}-repulsion that contributes to the reduction of the band gap.


For the GW corrections we use a model dielectric function for which only the high frequency dielectric constant $\epsilon_{\infty}$ has to be specified. For this parameter we have, for both polymorphs, used the experimental values $\epsilon_{\infty} = 12.3$ (InAs) and $\epsilon_{\infty} = 10.89$ (GaAs) \cite{Landolt1982}  corresponding to  the {\it zb} phase. The resulting quasiparticle band structures for {\it wz}-InAs are presented in Fig.~\ref{fig:GW_US_wz_val} and \ref{fig:GW_US_wz_core} (dashed lines) in the {\it d}-val and {\it d}-core case, respectively.
We cannot use calculated dielectric constants because they are dependent on the LDA gap and, hence, they are too large for both InAs and GaAs. In the InAs case, indeed, the $\Gamma_{1c} -  \Gamma_{15v}$ LDA gap is negative and, in the GaAs case, it is too small. Anyway, the dielectric constants do not differ substantially for the zinc-blende and the wurtzite phase, and this allow us to use the {\it zb} experimental value for both phases. The average value for wurtzite is, indeed, close to the zinc-blende value. This has been recently demonstrated for InN \cite{Furth2005} and ZnO \cite{Schleife2006}. The almost equal LDA gaps for zinc-blende and wurtzite InAs indicate a similar behavior.


 \subsection{The {\it pd} repulsion and the {\it d}-core case.}
 \label{3step_procedure}

The calculation of the band structure for the InAs compound is complicated by the presence of the {\it d} states, as discussed previously for systems like Ge,  II-VI semiconductors and some III-V semiconductors \cite{Furth2005, Tiago2004, Zakharov1994, Bachelet1985, Christensen1988, Wei1988}. Indeed, even tough the {\it d} states are well separated in energy from the lower lying {\it s} and {\it p} states with the same principal quantum number their wavefunctions have considerable spatial overlap which leads to a large exchange coupling between these states \cite{Rohlfing1995}. A proper description requires a pseudopotential which includes the {\it s} and {\it p} states in addition to the {\it d}  valence states. Such a description leads to expensive GW calculations. Another possibility is to use a pseudopotential which excludes the {\it d} states from the valence band.

Pseudopotential calculations based on the LDA and  treating the In 4{\it d} as valence states give smaller values of the gap due to the {\it pd} hybridizations of the upper valence band VBM ({\it p}-like states) with the {\it d} states. The strong {\it pd} interaction is mainly due to the fact that the LDA underestimates the binding energy of the {\it d} electrons. Consequently, the {\it d} levels are calculated to be too close to the top of the valence band and the {\it pd} coupling is overestimated within the LDA. This results in a shift of the {\it p} states towards higher energies and, hence, to the gap reduction.
The underbinding of the {\it d} states can be estimated by comparing the experimental ({\it zb}) and calculated positions of the {\it d} bands with respect to the VBM. 
The values calculated within the LDA are 14.3 eV and 14.5 eV for the {\it zb} and {\it wz} structures, respectively, compared to an experimental result of 16.8 eV \cite{Andersson1998} ({\it zb}).
Hence we obtain a scaling factor of 14.3/16.8 = 0.851 $\sim 85$\% for the {\it zb} and 14.5/16.8 = 0.857 $\sim 85$\% for the wurtzite. We will use this value of scaling factor to reduce the overestimation of the {\it pd}-repulsion calculated within the LDA.

\begin{figure}[htb]
\centering
\includegraphics[width=8cm]{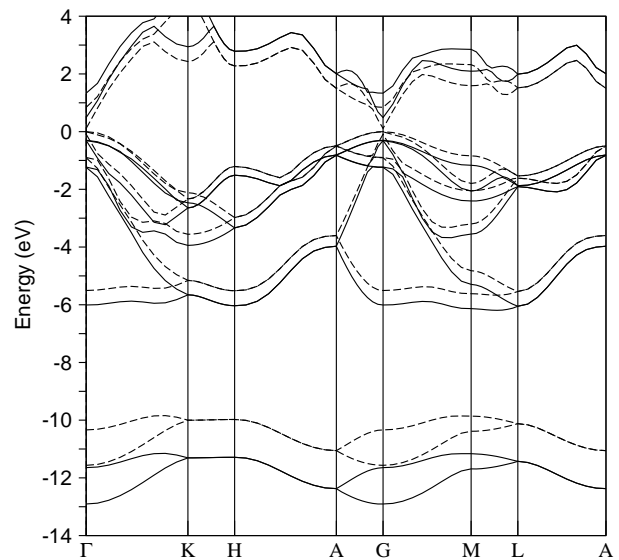}
\caption{Quasiparticle (solid lines) and Kohn-Sham (dashed lines) band structures of wurtzite InAs with the In 4{\it d} included among the core states.}
\label{fig:GW_US_wz_core} 
\end{figure}

In order to highlight the role played by the {\it d} states in the band structure calculations of these materials, we have studied both polymorphs with a (US) pseudopotential obtained by assuming the In 4{\it d} states to be core states while accounting for their influence by including self-interaction corrections (SICs) in the underlying calculations for the free atom \cite{Rieger1995}. Using this pseudopotential and the equilibrium lattice constant obtained in the {\it d}-val case, we have calculated the Kohn-Sham band structure and found that the correct band ordering is re-established. The semiconducting nature of InAs is now correctly predicted, with energy gaps being 
0.068~eV and  0.113~eV for the zinc-blende and the wurtzite phase, respectively.
Moreover, this calculations forms a good starting point for perturbatively adding GW corrections to the band structure in the usual way \cite{Hybertsen1986, Bechstedt1992b}. This is what we have done, {\it  i.e.} we have not attempted self-consistency with respect to the quasiparticle wave functions.
The so obtained LDA and GW band structures for the wurtzite polytype are reported in Fig.~\ref{fig:GW_US_wz_core}.

\begin{figure}[htb]
\centering
\includegraphics[width=8.5cm]{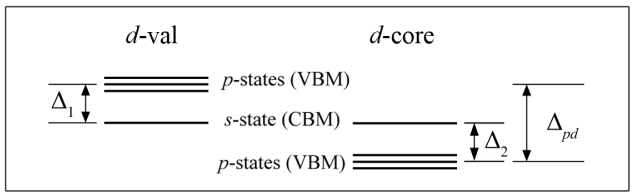}
\caption{Scheme for the estimation of the LDA value of the {\it pd}-repulsion ($\Delta_{pd}$).}
\label{fig:pd_repulsion} 
\end{figure}

The band structure obtained with the {\it d} electrons treated as core states misses the effects of the {\it pd} repulsion, {\it  i.e.},  the calculated GW band gap is larger than the ``true" one by the energy corresponding to the ``true"  {\it pd} repulsion itself.  Hence, we have here estimated this correction factor following the method outlined in \cite{Furth2005} in the case of InN. By comparing the LDA band structures in the {\it d}-core and {\it d}-val approaches, it is possible to estimate the LDA value for the  {\it pd} repulsion ($\Delta_{pd}$).  This can be identified with the shift of the {\it p} states toward higher energies when passing from the former 
to the latter
case. 
Indeed, when the {\it d} states are ``missing'' ({\it i.e.} very far below the VBM), there is no {\it pd} repulsion and it is possible to assume that the position of the CBM ({\it s}-like state) remains unchanged in the two situations. This is because in the InAs compound the {\it s-d} repulsion is negligible \cite{Persson2003}.  
Referring to the schematic representation in  Fig.~\ref{fig:pd_repulsion} and taking the  $\Delta_1$ and $\Delta_2$, as defined in this figure, from the calculated band structures, we obtain an estimate of $\Delta_{pd}$ for both polymorphs, that is 0.41~eV for the {\it zb} and 0.44~eV for the {\it wz} phase.
%

The {\it pd} repulsion obtained in this way, {\it i.e.} from  the LDA band structures, is overestimated with respect to the  ``true" one. To arrive at  the  ``true" value
$\Delta_{pd}$ must be reduced by the factor that relates the experimental energy difference between the VBM and the {\it d} levels to the same quantity within the LDA, {\it i.e.} by 85\% for both polymorphs. 
Hence the true {\it pd}-repulsion can be estimated as $\Delta_{pd} \times 0.85$, leading to  
 0.35 eV for the {\it zb} and 
0.38 eV for the {\it wz}. 
By applying this correction to the GW band structure calculated in the {\it d}-core approach we find that the quasiparticle energy gaps of  0.67~eV ({\it zb}) and 0.79~eV ({\it wz}) are reduced to 0.38~eV and 0.41~eV, respectively.


\section{LDA-SX-GW approach to the band structure calculation}
 
 \subsection{The InAs  case.}
 \label{LDA-SX-GW-InAs}

In the previous section we have obtained band gaps for the  two polymorphs of InAs using a method which is based on physical intuition but has some theoretical limitations.  It is based on experimental data for the {\it d}-band positions and can thus not be considered as fully {\it ab initio}. The method does not properly treat the In 4{\it d} states as valence states as one would  require from a proper theory. The advantage of the approach, however, is the remarkably low numerical effort and the fact that it succeeds in reducing the self-interaction among the very localized 4{\it d} electrons of In. 

We have, therefore, also applied a different approach to the problem, which is better defined from a theoretical point of view. The leading idea is to search for a starting point for the GW calculation which is closer to the quasiparticle electronic structure, thus approaching  a self-consistent GW calculation. As mentioned previously, in the {\it d}-val case  the LDA band structure is a bad starting point for calculating the GW corrections. Moreover, even if calculated correctly, it might be inadequate to add them in a perturbative way.

A full quasiparticle calculation in the GW approximation involves several difficulties such as the calculation of the frequency-dependent dielectric function and the solution of a Schr\"odinger-like equation with a non-local and energy dependent self-energy operator which is non-hermitian as opposed to the case of most eigenvalue problems involving a local energy independent potential. On top of these difficulties the calculations should actually be carried out self-consistently meaning that the output quasiparticle wave functions and energies should be used in calculating the self-energy operator from which the results are obtained.

Performing a self-consistent GW calculation is a heavy undertaking. 
It is well known that a non-selfconsistent calculation, {\it i.e.} approximating GW by ${\rm G_0W_0}$ where the screening (${\rm W_0}$) is taken from the random phase approximation (RPA), gives reasonable results.  The problem is that there is no systematic way of obtaining successively better approximations to the self-energy \cite{vonBarth1996} and, hence, for the self-consistency procedure. 
Indeed, when a full self-consistent GW calculation is performed, the self-consistent screening (W) is an auxiliary quantity with a rather unphysical behavior  \cite{Holm1998, Schone1998}.

In practice, one should always start the many-body perturbation expansion from a non-interacting Green function with one-electron eigenvalues close to quasi-particle energies. Then, self-consistency with respect to the eigenvalues can be carried out.
Self-consistency with respect to the wave functions raises a different problem: if the self-energy is energy dependent and non-Hermitian, the quasi-particle amplitudes are not orthogonal and cannot form the basis for a non-interacting Green function to start from. Hence, self-consistency with respect to wavefunctions should not be attempted. Indeed, when this was done, as in Ref. \cite{Popa2005}, the results were worse as compared to those obtained when the self-consistency was restricted to the eigenvalues.
Within the Screened-Exchange approximation (SXA), instead, the self-energy operator is Hermitian and energy independent. Hence, the wave functions are orthogonal and self-consistency can safely be performed.

Already the founder of the GW approach designed approximations to the GWA \cite{Hedin1965} in order to simplify its application  to real solids. One such approximation is the Coulomb-Hole plus Screened-Exchange approximation (COHSEX) which splits the self-energy of the GWA in two parts. The first part, the Coulomb-Hole term, was shown already by Hedin \cite{Hedin1965b} to have a rather weak dispersion although it is far from negligible. The main effect of this term is thus a constant shift of the entire band structure. Consequently, neglecting this term entirely and just keeping the Screened-Exchange (SX) term constitutes a rather good starting point for later adding the full GW correction in a perturbative way. Also, the use of the SX approximation is another way of achieving the drastic reduction of the self-interaction among the {\it d} electrons.

Viewing the exchange-correlation potential of the LDA as an approximation to the self-energy of the GWA and believing that the full GWA would predict InAs to be a semiconductor, the input LDA wave functions and energies are very far from the resulting wave functions and energies of the GWA. We here subscribe to the idea that the self-energy operator of the SXA being non-local but energy independent is much closer to that of the full GWA as compared to the LDA potential. 
The eigenvalues of the one-electron quasiparticle equation of the SXA are close enough to those of the GWA to allow us to obtain the full GWA results from first-order perturbation theory starting from the SXA.
This conjecture is strongly supported by recent results by some of us \cite{Fuchs2006}.
We thus use the wave functions and eigenvalues from the SXA calculation as input for the GWA calculation.

%

The calculations were performed by using the model dielectric function $\epsilon({\bf q}, \rho, \omega=0)$ of Ref. \cite{Bechstedt1992}. The dielectric constant $\epsilon_\infty$ of the material and the average electron density $\rho$ are required to build the static dielectric matrix. Since the LDA dielectric constant is too large due to the zero gap problem, we used the experimental value. The static dielectric matrix is then extended to finite frequencies by the generalized plasmon-pole model \cite{Hybertsen1986}.  

The one-electron part of our calculation is based on pseudopotentials obtained from the projector-augmented wave (PAW) method. This method allows for the reconstruction of the all-electron wave functions from their pseudo counterparts thus making it possible to calculate the important matrix elements of the SX operator between the valence states \cite{Wenzien1995,Furthmuller2002}, as well as to evaluate the nonlocal core-valence exchange \cite{Arnaud2000} .

%

We would like to stress that this is a very important advantage of using PAWs as compared to using ordinary norm-conserving or ultrasoft pseudopotentials. Unless some additional work toward estimating  the correct non-local core-valence exchange is done such ``normal" calculations can easily lead to errors in band gaps of the order of several tenths of an eV \cite{Kresse06}. The effect is more important in the case of valence $d$ states and must thus be accounted for in narrow-gap materials with strong $d$ contributions to states around the gap - as, e.g., in InAs. 
In the present work, the non-local core-valence exchange and correlation are approximated by pure exchange 
rather than by the dynamically screened potential of the GWA.  
We believe, however, that the effect of screening is rather small in the core region.

\begin{figure}[htb]
\centering
\includegraphics[width=8cm]{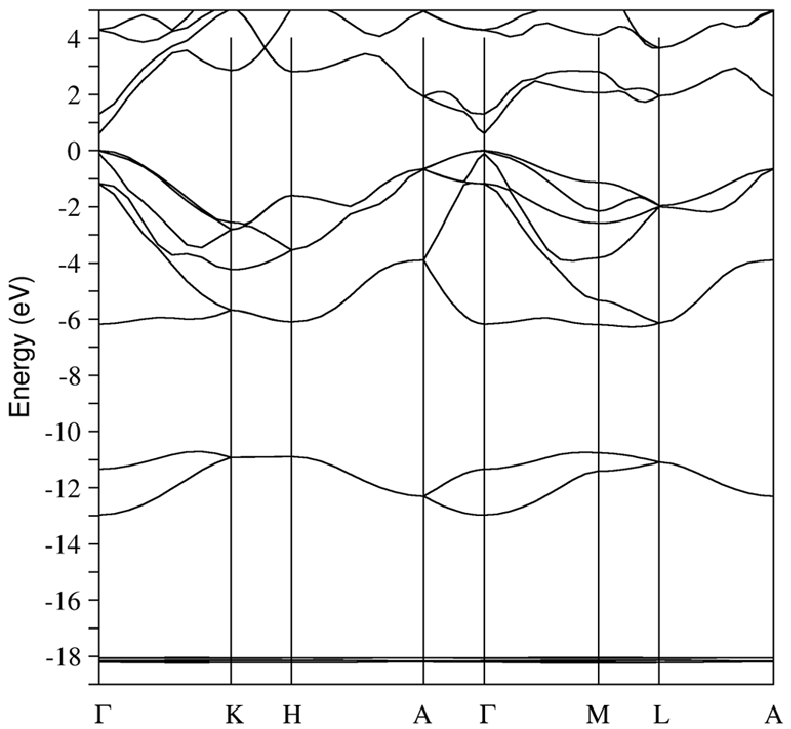}
\caption{GW band structure of InAs wurtzite with the In 4{\it d} included among the valence states. The quasiparticle corrections are applied on top of the SX calculation which, in turn, is started from the LDA eigenspectrum.}
\label{fig:GW514_wz_4d} 
\end{figure}

We find that the eigenenergies of the SXA calculation give the correct band ordering for both polymorphs, {\it i.e.}, positive band-gaps of 0.440 eV ({\it zb}) and 0.489  eV ({\it wz}).  
Consequently,  we  have a good starting point - the SX eigenspectrum - for the subsequent GW calculation and it is now possible to apply first-order perturbation theory.
We obtain a  quasiparticle spectrum with an energy gap of  0.556 eV for {\it zb} and 0.611 eV for {\it wz}.
The crystal field splitting for the {\it wz} phase obtained in this approximation amounts to  99 meV \cite{note1}.
The quasiparticle band structure of the InAs in the wurtzite phase with the In 4{\it d} included among the valence states is shown in Fig.~\ref{fig:GW514_wz_4d}. 

All the calculations are based on pseudopotentials obtained in the scalar-relativistic approximation. Hence, the calculated energy gaps should be compared with the experimental values modified as if there were no spin-orbit interaction. In the {\it zb} case the spin-orbit coupling splits the $\Gamma_{15}$ state by shifting the $\Gamma_{8v}$ states up by $+ \frac{1}{3}\Delta_0$ and the $\Gamma_{7v}$ state down by  $- \frac {2}{3}\Delta_0$. Since the {\it zb} gap at zero Kelvin is 0.415 eV and the spin-orbit splitting $\Delta_0 = 0.38$~eV \cite{Landolt1982}, it follows that the experimental energy gap of the zinc-blende in a scalar-relativistic world would be 
0.542 eV. This is the value which should be compared to our calculation.  In Table~\ref{gaps} we see that our result compares well with experiment
and that this approach leads to a better agreement than the three step procedure of section \ref{3step_procedure}.

Encouraged by the good results for the zinc-blende structure, we can estimate what the wurtzite energy gap would be in the real world. Indeed the difference between the two polymorphs is mainly in the stacking of the atoms (A B C sequence for the {\it zb} and A B A B for the {\it wz}), that is, they differ in the second nearest neighbor of a given atom. Hence, it would not be too far fetched to assume that the errors introduced by our approximations will be similar in the two cases, thereby allowing us to estimate the change in the energy gap due to the change in the crystal structure as the difference between the calculated {\it wz} and {\it zb} values, {\it i.e.}

\begin{equation}
E_{gap}(wz) - E_{gap}(zb) = 55 ~{\rm meV}.  
\end{equation}

By adding this energy to the experimental gap (at 0~K) of InAs in the zinc-blende phase we obtain the following estimate of the gap (at 0~K) of  the wurtzite polymorph 

\begin{equation}
E_{gap}(wz)  =  470 ~ {\rm meV}.   
\end{equation}

Results from low temperature (5 K) photocurrent measurements on thick ($\sim 85~$nm in diameter) InAs-InAsP-InAs nanowires of Ref \cite{Tragardh2006}  suggest a value of $\sim 550$~meV for the energy gap of InAs in the wurtzite phase, in good agreement with our result. Using this value and by assuming that the spin-orbit splitting for the {\it wz} polymorph is similar to that of the {\it zb}, we find the experimental gap of InAs in the {\it wz} phase without spin-orbit interaction to be  0.677 ~eV. 

\begin{table}[thbp]
\begin{center}
    \begin{tabular}{@{} cccccccc @{}}
    \toprule
    
      				 	                   
 poly  & $\diagdown E{_{gap}} $  &   LDA         &   LDA          & GW              & SX              &  GW    &  EXP \\ 
 type  &                                        &  {\it d}-val   & {\it d}-core  & {\it d}-core   & {\it d}-val     &  {\it d}-val  &   w.out S-O  \\
  \hline
    {\it zb}          	&			& $-0.346$      &  0.068           &  0.38    	      & 0.440    &  0.556                     &      0.542   \\ 
    {\it wz}     	&			&  $-0.317$     &  0.113           &   0.41           & 0.489    &  0.611                     &      0.677     \\ 
  \hline \hline
  \end{tabular}
\end{center}
\caption{InAs energy gaps at ${\Gamma}$ point calculated in the different approximations outlined in the article:
(i) DFT-LDA with In 4{\it d} included among the valence states ({\it d}-val)
(ii) DFT-LDA with In 4{\it d} as core states described by SIC pseudopotentials ({\it d}-core)
(iii) GW for the {\it d}-core case after the correction for the {\it pd}-repulsion
(iv) SX calculation with PAW pseudopotential in the  {\it d}-val case
(v) GW gap resulting from the LDA-SX-GW method with PAW pseudopotential in the  {\it d}-val case
(vi) experimental gap without spin-orbit interaction. 
All values are given in eV.
}
\label{gaps}
\end{table}

\subsection{The GaAs  case.}

In analogy to the InAs case, we use PAW pseudopotentials which treat the Ga 3{\it d} electrons as valence states. Even though GaAs is correctly predicted to be a semiconductor by the LDA, those band gaps are quite small compared to experiment, as can be seen from Table~\ref{GaAs-gaps}. When we compute the quasiparticle corrections starting from these LDA  eigenfuctions, we find an energy gap which is still too small - they are of the order of $\sim 1$~eV in the {\it zb} case. A similar result is also reported in the case of all-electron GW calculations \cite{Arnaud2000}, indicating that one needs to go beyond the non-selfconsistent GWA in order to better describe the GaAs band structure.

\begin{figure}[htb]
\centering
\includegraphics[width=8cm]{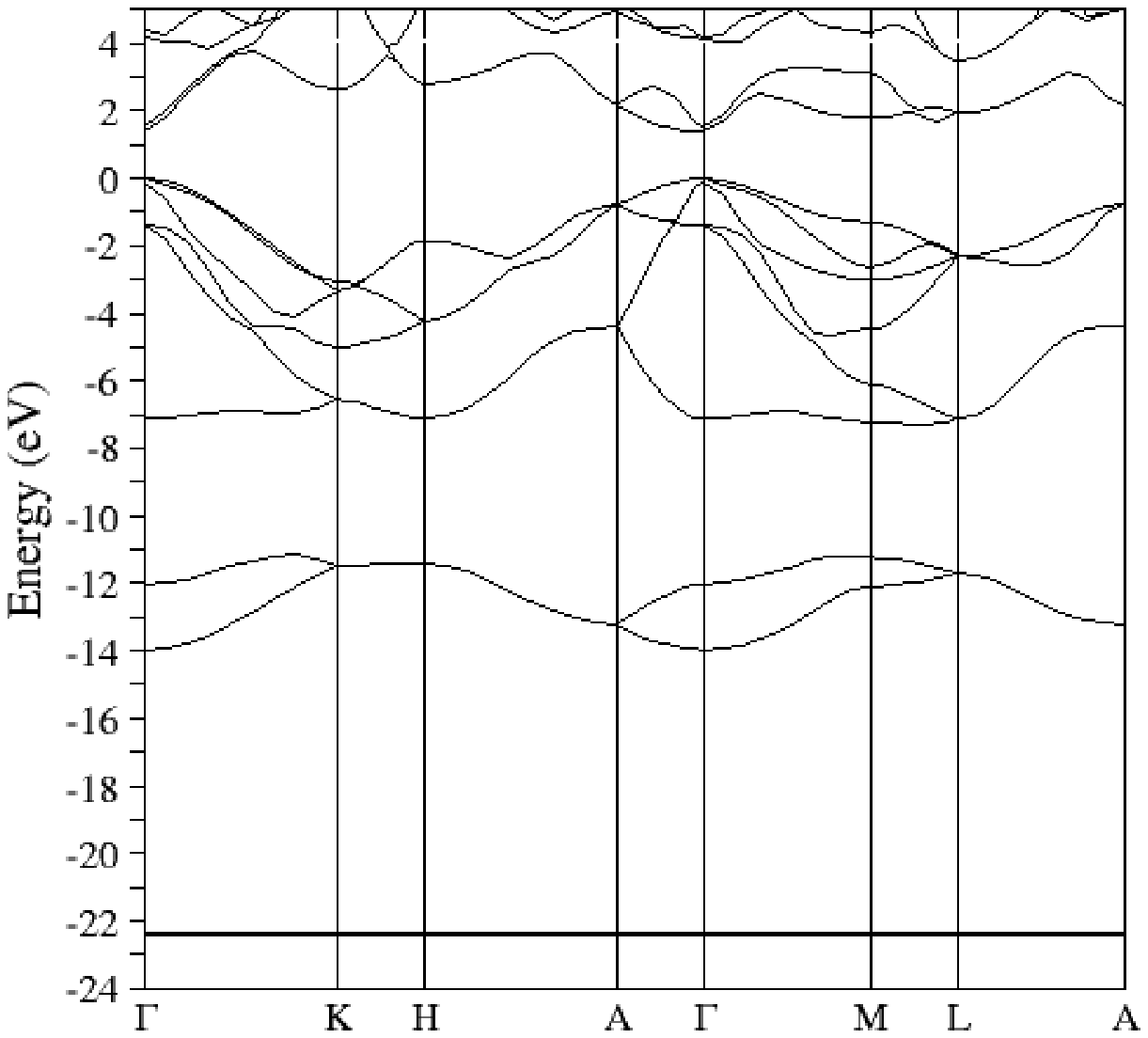}
\caption{Band structure of GaAs wurtzite with Ga 3{\it d} included among valence states. The quasiparticle corrections are applied on top of the SX calculation which, in turn, is started from the LDA eigenspectrum.}
\label{fig:GaAs-wz} 
\end{figure}

We have, thus, applied the procedure described in Section \ref{LDA-SX-GW-InAs} also to GaAs. The resulting  energy gaps are summarized in Table~\ref{GaAs-gaps} and the full quasiparticle band-structure is reported in Fig.~\ref{fig:GaAs-wz} in the {\it wz} case. From these results we note that, at the LDA level of the calculation, the gap of the {\it wz} polymorph is $50$~meV larger than the {\it zb} case, consistent with Ref. \cite{Yeh1994} and the result for InAs.  When the SX calculation is performed this behavior is reversed and the {\it zb} gap is larger than the {\it wz}.  This  situation again changes after the final GW calculation, with the {\it zb} gap being smaller by $219$~meV. These calculations  were performed using the same $k$-point mesh as for the InAs case \cite{note1}.

It is well known that going from a local potential like, e.g., the LDA potential to a non-local and energy independent potential like that of the Hartree-Fock approximation (HFA) has a drastic effect on band gaps. The SX approximation is less extreme than the HFA but we still expect a substantial effect on the band gap. This is what we actually observe in both semiconductors: in the GaAs case the SXA results in a {\it zb} gap larger then the {\it wz} one and, in the InAs case, the SXA calculation changes the wrong ordering of the bands into the correct one.

Besides, we note that there is a delicate balance that may be changed in the various approximations of exchange and correlation LDA, SX, GW used. GaAs is more sensitive than InAs in this respect. Empirically one observes that with increasing size difference between cation and anion and increasing ionicity of the bonds the wurtzite gap is increased with respect to the zinc-blende gap. For GaAs, however, the size difference vanishes and the ionicity (compared with ZnO, InN) is the smallest.

Photoluminescence measurements were recently performed on single GaAs nanowires using the NSOM technique \cite{Mintairov}. These NWs have mixed crystal structure:  segments having {\it zb} and {\it wz} crystal structures can be identified in the same wire, as reported in \cite{Soshnikov}. Since these wires have large {\it wz} segments, it was possible to take the PL emission from that part of the wire finding that it has emission energy lower than that of the {\it zb} by $\sim 50$~meV, {\it i.e.} the measured energy gap of the GaAs in the {\it wz} phase at low temperature ($\sim 10$~K) was found to be 1.467~eV. Since these NWs are typically 50 nm in diameter, we can neglect quantum confinement effects.
Our GW result is in contrast with these measurements, nevertheless we would like to suggest  more experiments on GaAs in the wurtzite phase, especially experiments performed on purely wurtzite GaAs NWs.

To compare our calculated gaps with experimental data we strip the experimental results from their spin-orbit contribution, using the same procedure as in the InAs case. By using  the value 1.519~eV and $\Delta_0 = 0.33$~eV for the low temperature energy gap and spin-orbit splitting in GaAs ({\it zb}) \cite{Landolt1982} we find that the energy gap without spin-orbit interaction is 1.629 eV. Assuming that the GaAs in the {\it wz} phase has a spin-orbit splitting similar to the {\it zb} one and using the results from Ref. \cite{Mintairov}, we found an  experimental {\it wz} gap ``without" spin-orbit interaction of 1.577~eV.

      
\begin{table}[thbp]
\begin{center}
    \begin{tabular}{@{} ccccccc @{}}
    \toprule
     poly & $\diagdown E{_{gap}} $  &   LDA     &   SX      & GW         &  EXP   &  EXP \\  
     type &                                                &              &               &               &            &  w.out S-O \\
  \hline
    {\it zb}  &   & $0.330$       &  1.289      	 &   1.133          &      1.519		  &  1.629   \\ 
    {\it wz}  &  &  $0.380 $  	  &  1.172          &   1.351          &      1.467                 &  1.577      \\ 
  \hline \hline
  \end{tabular}
\end{center}
\caption{GaAs energy gaps at ${\Gamma}$ point calculated according to different approximations: LDA (i), SX (ii) and LDA-SX-GW (iii). Experimental gap for the {\it wz} \cite{Mintairov} and the {\it zb} with (iv) and without (v) spin-orbit interaction. All the calculated band gaps are obtained using PAW pseudopotentials and including the Ga 3{\it d} among the valence states. All the values are in eV.}
\label{GaAs-gaps}
\end{table}

\section{SUMMARY AND CONCLUSIONS}

We have reported on the first study of the quasiparticle band structure of InAs and GaAs in the wurtzite phase. The calculations were performed within an approximately self-consistent GW approach. 
For the purpose of comparison we have done this in the InAs case also by a three-step procedure  based on the LDA in which the In 4{\it d} electrons were frozen and treated as core electrons. 
This resulted in a band structure with the correct band ordering. We then calculated the GW corrections to this results in a perturbative manner and added a correction designed  to account for the missing {\it pd}-repulsion. The latter physically rather intuitive approach led to results not to far from those of the more fundamental approach, which in turn was in good agreement  with recent experimental results. This demonstrates the importance of self-consistency in these materials where the LDA results are  quite off the mark and actually predict InAs to be a metal.

We have found that the InAs wurtzite energy gap is larger than the zinc-blende one by $\sim 55$~meV, leading to the theoretical estimate of the quasiparticle gap of InAs in the wurtzite phase ($\sim 0.47$~eV), a value in close agreement with very recent measurements on InAs based nanowires.

The quasiparticle energy gap of GaAs in the {\it wz} phase is also larger than that of the {\it zb} polytype. This finding is, however, in disagreement with preliminary photoluminescence experiment performed on GaAs nanowires having a wurtzite crystal structure.



\begin{acknowledgments}
We thank the Nanoquanta network of Excellence (contract number NMP4-CT-2004-500198), the Deutsche Forschungsgemeinschaft (Project
No. Be 1346/18-1) and the European CommunityÕs Human Potential Program in the  framework of the  Photon-Mediated Phenomena (PMP) Research Training Network under contract HPRN-CT-2002-00298 for supporting this research.
\end{acknowledgments}

%


\end{document}